\documentclass[twocolumn,american,prl,showpacs,superscriptaddress]{revtex4-1}
\usepackage{amsmath} 
\usepackage{graphicx} 
\usepackage{amssymb}
\usepackage{times} 
\usepackage{color} 
\usepackage{comment}

\usepackage[abs]{overpic}
\usepackage{subfigure}
\usepackage[dvipsnames,svgnames]{xcolor}

\definecolor{lr}{rgb}{1.0,0.3,0.3} \definecolor{dg}{rgb}{0.0,0.5,0.0}



\begin{document}

\title{\emph{Ab initio} calculation of spin-orbit coupling for NV center in
diamond exhibiting dynamic Jahn-Teller effect}

\author{Gerg\H{o} Thiering} \affiliation{Wigner Research Centre for Physics,
Hungarian Academy of Sciences, PO Box 49, H-1525, Budapest, Hungary}
  
 
\author{Adam Gali} \email{gali.adam@wigner.mta.hu} \affiliation{Wigner Research
Centre for Physics, Hungarian Academy of Sciences, PO Box 49, H-1525, Budapest,
Hungary} \affiliation{Department of Atomic Physics, Budapest University of
Technology and Economics, Budafoki \'ut 8., H-1111 Budapest, Hungary}
\email{gali.adam@wigner.mta.hu}


\begin{abstract} Point defects in solids may realize solid state quantum bits.
The spin-orbit coupling in these point defects plays a key role in the
magneto-optical properties that determine the conditions of quantum bit
operation. However, experimental data and methods do not directly yield this
highly important data, particularly, for such complex systems where dynamic
Jahn-Teller (DJT) effect damps the spin-orbit interaction. Here, we show for an
exemplary quantum bit, nitrogen-vacancy (NV) center in diamond, that \emph{ab
initio} supercell density functional theory provide quantitative prediction for the
spin-orbit coupling damped by DJT. We show that DJT is responsible for the multiple intersystem crossing rates of NV center at cryogenic temperatures. Our results pave the way toward optimizing solid state
quantum bits for quantum information processing and metrology applications.
\end{abstract} \maketitle



\begin{figure}
	\includegraphics[width=1.00\columnwidth]{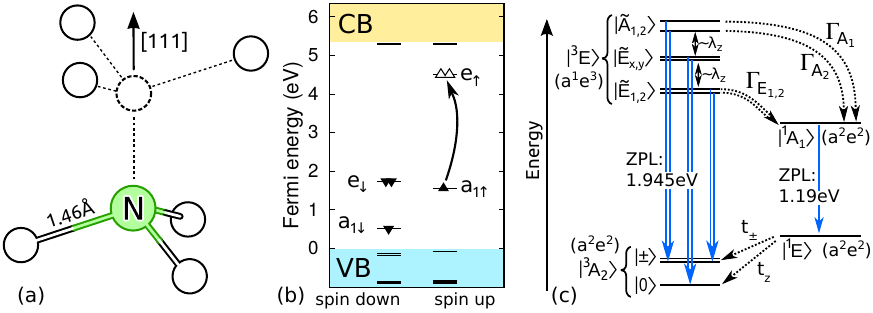}
\caption{\label{fig:NV}NV center in diamond. (a) Schematic diagram of the structure of the negatively charged defect with the optimized carbon-nitrogen bond length. The symmetry axis of the defect in the diamond lattice is shown. (b) The calculated defect levels in the gap are depicted in the ground state where the curved arrow symbolizes the $\Delta$SCF procedure for creating the triplet excited state. The $e$ states are double degenerate. VB and CB corresponds to valence and conduction bands, respectively. (c) The corresponding ground state and excited states are shown as well as the optical electron
spin polarization cycle. The spin-orbit splitting $\lambda_z$ is depicted that separates the sublevels in the triplet $^3E$ excited state. The corresponding intersystem crossing rates between the $^3E$ substates ($\tilde{A}_{1,2}$, $\tilde{E}_{1,2}$ double group representations) and the singlet $^1A_1$ is labeled by $\Gamma$s. The tilde labels the vibronic nature of these states. The intersystem crossing ($t_\pm$ and $t_z$) from the $^1E$ to the triplet ground state is shown for the sake of completeness that closes the spin-polarization cycle.} 
\end{figure}

Dopants in solids are promising candidates for implementations of quantum bits
for quantum computing~\cite{Ladd:Nature2010, Awschalom2013}. In particular, the
negatively charged nitrogen-vacancy defect (NV center) in
diamond~\cite{duPreez:1965} has become a leading contender in solid-state
quantum information processing because of its long spin coherence time (up to several milliseconds in ultra-pure diamonds~\cite{Balasubramanian:NatMat2009}) and ease of optical initialization and readout of spin state,\cite{Jelezko:PSSa2006} even non-destructively~\cite{Awschalom:Nature2010,Robledo:Nature2011}. NV center consists of a nitrogen atom
substituting a carbon atom in the diamond lattice adjacent to a vacancy with
forming a $C_{3v}$ symmetry that dictates the selection rules for the
magneto-optical processes [Fig.~\ref{fig:NV}(a)]. The geometry and the electronic structure of NV center in diamond have been discussed previously based on highly converged \emph{ab initio} plane wave large supercell calculations~\cite{Gali:PRB2008, Gali2009}. The defect
exhibits a fully occupied lower $a_1$ level and a double degenerate upper $e$ level filled by two parallel-spin electrons in the gap with comprising an $S=1$ high-spin ground state [Fig.~\ref{fig:NV}(b)]. The high-spin $^3E$ excited state can be well-described by promoting an electron from the lower defect level to the upper level in the gap~\cite{Gali:PRL2009}. Between excited state and ground state triplets dark singlets appear that can selectively flip $m_s=\pm 1$ states to
$m_s=0$ state in the optical excitation cycle \cite{Harrison:DRM2006,
Wrachtrup:JPCM2006, Manson:PRB2006, Maze:NJP2011, Doherty:NJP2011}, i.e., the electron
spin can be spin-polarized optically [Fig.~\ref{fig:NV}(c)].
The spin-selective decay is
predominantly caused by the intersystem crossing (ISC) between the optically
accessible triplet $^3E$ excited state [see Fig.~\ref{fig:NV}(c)] and the singlet $^1A_1$ state which is
mediated by the spin-orbit coupling between these states and
phonons~\cite{Maze:NJP2011,Doherty:NJP2011}. Generally, the spin-orbit coupling
plays a crucial role in the optical spin-polarization and readout of NV quantum
bit (qubit) and alike \cite{Koehl11, Soltamov2012SiV_sic, Hepp:2014SiV, Nadolinny2016epr, christle:2017}.

The strength of the spin-orbit coupling in NV center may be detected at low
temperature photoluminescence excitation (PLE) measurements of the $^3E$
excited state in high quality diamond samples \cite{Batalov:PRL2009} where the
strain
does not deteriorate the fine spin level structure of the $^3E$ excited state
\cite{Maze:NJP2011}. By combining the PLE data and group theory analysis, the
spin-orbit strength was estimated to be 5.33$\pm$0.03~GHz~\cite{Batalov:PRL2009,
Bassett:2014Science} recorded at $T<20$~K temperatures. We note that  additional splitting and shift of the levels appear due to electron-spin--electron-spin 
interaction~\cite{Maze:NJP2011} that is not discussed further here.  Regarding the spin-orbit mediated ISC process, three different ISC rates have been recently deduced in the experiments \cite{DohertyGoldman2015}. This is surprising because group theory \cite{Maze:NJP2011} predicts only a single decay channel from the $A_1$ substate of the $^3E$ triplet. This phenomenon was qualitatively explained by assuming weak coupling of symmetry breaking acoustic $e$ phonons to the $A_1$, $E_{1,2}$ and $A_2$ substates that was derived in the frame of perturbation theory of electron-phonon coupling \cite{DohertyGoldman2015, Goldman:PRB2015}. Nevertheless, this theory does not quantitatively account on the experimental ratio between the ISC rates at cryogenic temperature.  
 
Deep understanding of the nature of $^3E$ excited state might resolve this issue. \emph{Ab initio} simulations \cite{Abtew:PRL2011} and
experiments~\cite{Fu:PRL2009, PlakhotnikManson:PRB2015, Ulbricht:2016Ncomm} imply that the $^3E$ excited state exhibits
dynamic Jahn-Teller (DJT) effect where the vibronic levels split in 10~meV
region~\cite{Abtew:PRL2011}. This is more than three orders of magnitude larger
than the observed spin-orbit coupling, thus spin-orbit coupling can be
considered as a perturbation with respect to the DJT effect. In DJT systems, the phonons and orbitals are strongly coupled that goes \emph{beyond} perturbation theory of electron-phonon coupling. That may lead to the
damping or even quenching of spin-orbit coupling~\cite{Ham:PR1965} where the
damping parameter, i.e, the Ham reduction factor depends on the strength of electron-phonon coupling. For this
reason, the strength of the intrinsic spin-orbit coupling in NV center is still an open
question. In addition, the presumably strong electron-phonon coupling in $^3E$ state may have implications in the ISC process of NV center.

Here we show by \emph{ab initio} supercell density functional theory (DFT)
calculations that the intrinsic spin-orbit coupling together with the damping caused
by DJT can be accurately predicted for NV center in diamond. We show that strong electron-phonon coupling in the triplet excited state is an important aspect of the theory of non-radiative decay in NV center, and this novel theory accurately reproduces the ratio between the experimental ISC rates.

\label{sec:method}

We apply supercell plane wave spin-polarized density functional theory (DFT)
method to model
the defect with the usual Born-Oppenheimer approximation. The Born-Oppenheimer 
approximation treats the nuclei or ions as classical particles, where the motion
of the electrons
are solved in fixed coordinates of these ions providing the total energy of the
system parametrized
by the coordinates of ions. 
By varying the coordinates of ions one can achieve an adiabatic potential energy
surface (APES)
map of the system. The global minimum of APES defines the optimized geometry of
the system.
The NV center was modeled mostly in a 512-atom cubic supercell within
$\Gamma$-point
approximation that provides convergent electron charge density. In our
simulations, we apply \textsc{VASP~5.4.1} plane wave code~\cite{Kresse:PRB1996}
within
projector-augmentation-wave-method (PAW)~\cite{Blochl:PRB1994,Blochl:PRB2000} to
treat
the ions. We utilize
the standard PAW-potentials and a convergent plane wave cutoff of 370~eV. We
apply a very stringent upper limit of 10$^{-4}$~eV/\AA\ for the forces in the
geometry optimization. For the calculation of the phonon spectrum we generate
the dynamical matrix via finite differences of total energies. We applied the Huang-Rhys theory to calculate the overlap of phonon states \cite{Alkauskas:NJP2014, SupplMat}. The implementation is described in our previous publication \cite{Gali:2016Ncomm}. The $^3E$ state is calculated by constrained-occupation DFT method by following the picture in Fig.~\ref{fig:NV}(b)
($\Delta$SCF)~\cite{Gali:PRL2009}. The singlet states cannot be described properly by conventional Kohn-Sham density functional theory because of the high correlation between the $e$ orbitals. A crude estimation of the $^1A_1$ state is to apply spinpolarized singlet occupation of $e_x$ orbital. We use this method to estimate the geometry of  $^1A_1$ needed in the description of ISC.
The optimized geometry in the excited state is calculated by minimizing the forces
acting on the atoms in the electronic excited state within $\Delta$SCF
procedure. The electronic structure is calculated using HSE06 hybrid functional
\cite{Heyd03, Krukau06} within DFT. Using this technique, it is possible, in
particular, to reproduce the experimental band gap and the charge transition
levels in Group-IV semiconductors within 0.1~eV accuracy \cite{Deak:PRB2010}.

The spin-orbit coupling (SOC) was calculated within non-collinear approach as
implemented
in \textsc{VASP~5.4.1} (see ref.~\onlinecite{Steiner:2016PRB}). 
In our particular cases, the $C_3$ axis of the defect set the quantization
axis of the spin that fixed in the calculation. As SOC is a tiny perturbation to
the system we also
fixed the coordinates in SOC calculations as obtained from the spin-polarized
DFT calculations.
Direct comparison to the experimental data can be extracted from the $z$
component of SOC
which coincides with the quantization axis of the spin, thus it is the diagonal
term in the matrix
elements representing the different components of SOC. The spin-orbit scattering
rate, however, can be calculated from the $\lambda_\perp$ components that 
are off-diagonal terms in the SOC Hamiltonian. SOC was calculated in the
high
symmetry configurations of NV($-$) center. 
In this case, a single electron should occupy the double degenerate Kohn-Sham
$e$ state in the spin minority
channel. After applying SOC this double degenerate Kohn-Sham state will split to
$e_x$ and
$e_y$ states. By using $\Delta$SCF procedure it is feasible either to occupy the $e_x$
or $e_y$ state,
and the calculated total energy difference is the strength of SOC. We find that the
half of this total energy difference is equal within
$10^{-7}$~eV to the split of $e_x$ and $e_y$ Kohn-Sham levels when these two states are occupied by half-half electrons. Thus, the strength of SOC can be calculated by the
half-half occupation of the
$e$ states with following the SOC splitting of these $e$ states. It is crucial to use $\Gamma$-point approximation in the integration of the Brillouin-zone in this procedure because the degenerate $e$ orbitals slightly split in other $k$-points by reducing the symmetry of the orbitals that makes the readout of the SOC parameters ambiguous. 
We further note that $\left\langle^3E (A_1) | H_\text{SO} | ^3E (A_1)\right\rangle = \frac{1}{2}\left\langle e_{+} | H_\text{SO} | e_{+}\right\rangle$ and  $\left\langle^3E | H_\text{SO} | ^1A_1\right\rangle = \frac{1}{\sqrt{2}}\left\langle e_{+}^\downarrow | H_\text{SO} | a_{1}^\uparrow\right\rangle$ for NV($-$) defect where the latter is valid with the two basic assumptions described below in the discussion of ISC rates. Here the arrows represent the corresponding spin states and $\left|e_{\pm}\right\rangle= \frac{1}{\sqrt{2}}(\left|e_x\right\rangle \pm i \left|e_y\right\rangle)$. We note that we calculated SOC for the considered defects by Perdew-Burke-Ernzerhof (PBE) DFT functional \cite{PBE} too for test purposes that is a significantly faster method than the HSE06 DFT method. In that case we used PBE optimized lattice constants and defect geometries.

%
%


First, we consider the spin-orbit splitting $\lambda_z$.  Batalov and co-workers deduced a value of 5.3~GHz for this spin-orbit coupling (SOC) from PLE measurements~\cite{Batalov:PRL2009} observed at 6~K temperature. We study this interaction by our \emph{ab initio} method. The convergent SOC was achieved with scaling supercell size up to 1000-atom in the hybrid functional calculations and apply an exponential fit to the calculated results (see Ref.~\onlinecite{SupplMat}). 
The calculated $\lambda_z$ for the $^3E$ state of NV($-$) is 15.8~GHz which is about $3\times$ larger than the measured one. However, the $^3E$ excited state is principally Jahn-Teller unstable in the high $C_{3v}$ symmetry that can lead to partial or full quenching of the spin-orbit coupling~\cite{Ham:PR1965}.

\begin{figure}
	\includegraphics[width=0.7\columnwidth]{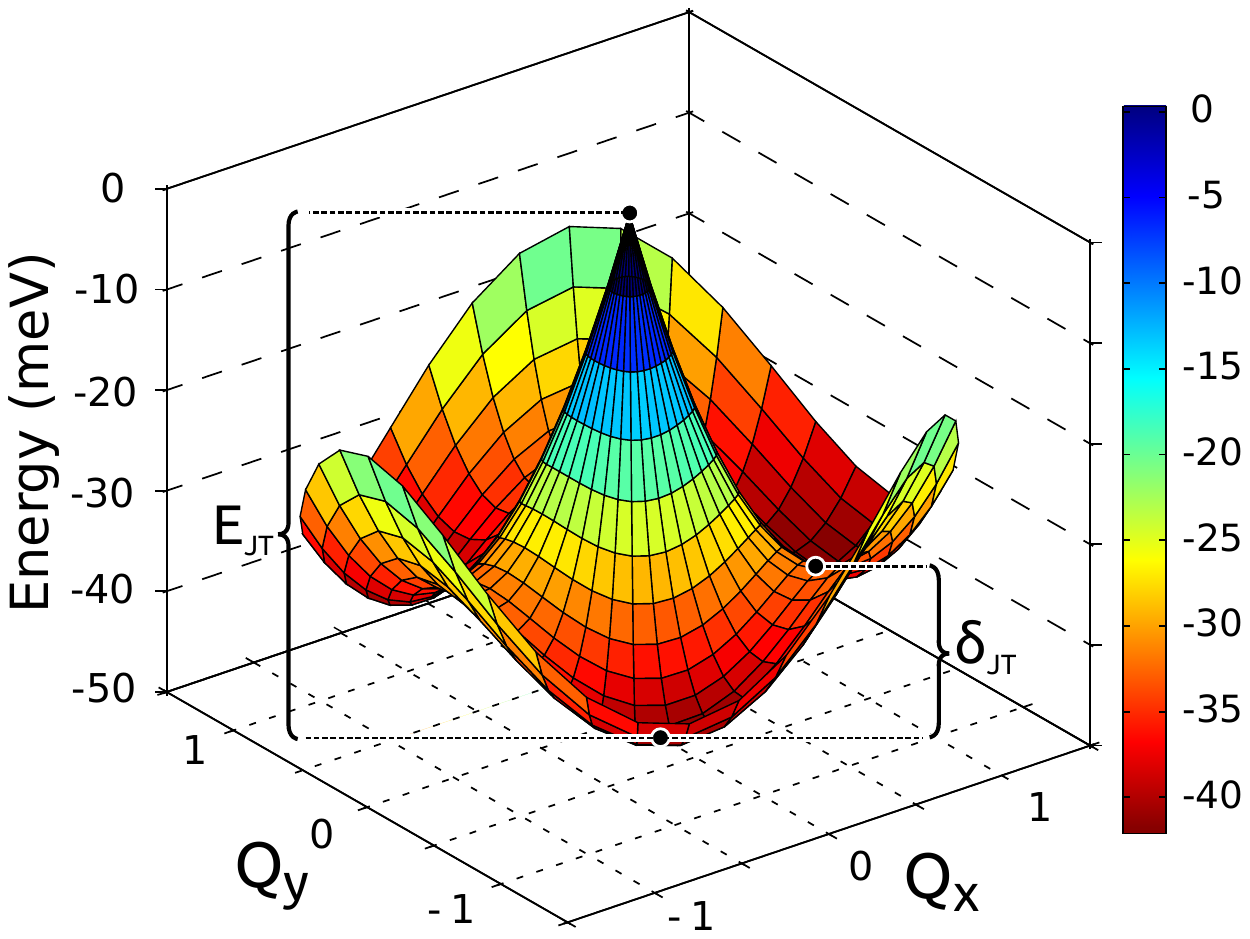}
	\caption{\label{fig:APES}Adiabatic potential surface (APES) of the quadratic DJT system for NV($-$) $^3E$ excited state. $Q_{x,y}$ configuration coordinates represent the degenerate $e$ phonon. $E_\text{JT}$ is the Jahn-Teller energy which is the energy difference between the total energy at the high symmetry configuration and the distorted configuration.  The energy barrier of $\delta_\text{JT}$ occurs between the three equivalent distorted configurations.}
\end{figure}

We conclude that the spin-orbit coupling should be studied \emph{beyond} Born-Oppenheimer approximation. $^3E$ state is orbitally degenerate where a symmetry breaking $e$ phonon or quasi-local vibration mode can drive out the system from the high symmetry that may couple the components of the double degenerate $E$ electron wavefunctions. This is a so-called $E\otimes e$ DJT system that was already analyzed in detail by Ham~\cite{Ham:PR1968} and
Bersuker~\cite{bersuker2013jahn,bersuker2012vibronic}. Here,  Since the defect has $C_{3v}$ symmetry thus the quadratic JT Hamiltonian should be considered (see the analysis in Ref.~\onlinecite{SupplMat}). By introducing the dimensionless coordinates along $x=a^\dag_x+a_x$ and $y=a^\dag_y+a_y$, the quadratic DJT Hamiltonian reads as
\begin{widetext}
\begin{equation}
\label{eq:Exe_quad}
\hat{H}=\hbar\omega_{e}\left(a_x^\dag a_x +a_y^\dag a_y+1\right)+F\left(x\sigma_{z}+y\sigma_{x}\right)+G\left[\left(x^{2}-y^{2}\right)\sigma_{z}+2xy\sigma_{x}\right] \text{,}
\end{equation}
\end{widetext}
where $a_{x,y}$, $a^{\dagger}_{x,y}$ operators create or annihilate an $e$
vibration. The double degenerate $e$ mode has $x$ and $y$ components. $\hbar \omega_e$ is the energy of the effective $e$
mode that
drives the distortion of the system, and $F$ and $G$ are electron-vibration coupling related terms.
 The electrons are
 represented by the
Pauli matrices $\sigma$. $F$ and $G$ can be directly derived from the calculated APES which results in the Jahn-Teller energy ($E_\text{JT}$) and the barrier energy $\delta_{JT}$ between the global minima (see Fig.~\ref{fig:APES}) as follows: $E_\text{JT}=\frac{F^2}{2\hbar \omega_e}$, $G=\delta_{JT} \hbar \omega_e / 2 E_{JT}$. The $\hbar \omega_e$ energy can be derived from the parabola fitting to the calculated APES. Finally, all the parameters can be readout from APES (see Table~\ref{tab:results}) that allows to solve Eq.~\eqref{eq:Exe_quad} numerically that provides the electron-phonon coupling coefficients \cite{SupplMat}. The exact solution can be expanded into series as
$\left|\Psi_{\pm}\right\rangle =\sum_{nm}\left[c_{nm}\left|E_{\pm}\right\rangle \otimes\left|n,m\right\rangle +d_{nm}\left|E_{\mp}\right\rangle \otimes\left|n,m\right\rangle \right]$,
where we limit the expansion up to four oscillator quanta ($n+m\leq4$) which is numerically convergent within 0.2\%. The $p$ reduction factor that reduce the spin-orbit interaction can be then calculated from these coupling coefficients as $p=\sum_{nm}\left[c_{nm}^{2}-d_{nm}^{2}\right]$ which represents the mixture of
the $E^+$ component with the $E^-$ component of the $^3E$ state that results in  the quenching of the effective angular momentum.  We note that by taking only the linear term either numerically or approximately~\cite{Ham:PR1968} results in 10\% lower value for the damping factor (see details in Ref.~\onlinecite{SupplMat}).

The calculated quenching factor is large that can strongly modify the intrinsic (purely orbital) value. The final result is 4.8~GHz that agrees within 10\% with the experimental result (see Table~\ref{tab:results}). Despite the remaining discrepancy between the calculated and measured SOC, the improvement is giant in SOC when DJT effect is considered that demonstrates the complex physics
of this system. We emphasize that this result was obtained from first principles 
calculations in the microelectronvolt energy region of SOC.
\begin{table}[htbp]
	\caption{SOC splitting parameters for the excited state of NV($-$).  By using $E_\text{JT}$ Jahn-Teller energy, $\hbar\omega_e$ phonon energy, and $\delta_\text{JT}$ barrier energy the $p$ reduction factors can be evaluated. The partially quenched  $p \lambda_z$ SOC parameters are expected to be observed in the experiments. The calculated values are valid at 0~K temperature. We note that PBE functional results in $E_\text{JT}=25$~meV and $\delta_\text{JT}=10$~meV (see Ref.~\onlinecite{Abtew:PRL2011})}
	\begin{ruledtabular}
		\begin{tabular}{lllllll}        
			 $\lambda_z$ & $E_\text{JT}$  &  $\delta_\text{JT}$ & $\hbar\omega_e$  & $p$ & $p \cdot \lambda_z$  & exp. \\
			(GHz) & (meV) & (meV) & (meV) & & (GHz) & (GHz) \\
			\hline
			    15.8
			&  41.8
			&  9.1
			&  77.6
			&  0.304
			& 4.8 
			&  5.3\footnote{$\lambda_z$ in ref.~\onlinecite{Batalov:PRL2009} measured at 6~K temperature.}, 5.33$\pm$0.03\footnote{$\lambda_z$ in ref.~\onlinecite{Bassett:2014Science} observed below 20~K temperatures.} 
		\end{tabular} 
	\end{ruledtabular}     
	\label{tab:results}
\end{table}

We turn to the investigation of the ISC process which depends on the perpendicular component of the spin-orbit coupling, $\lambda_\perp$ \cite{Maze:NJP2011, Doherty:NJP2011}. The ISC rate between the triplet $^3E$ and $^1A_1$ may be calculated \cite{DohertyGoldman2015} as 
\begin{equation}
\label{eq:ISCrate}
\Gamma_{A_1} = 4 \pi \hbar \lambda_\perp^2 F(\Delta) \text{,}
\end{equation}
where $\lambda_\perp$ is given in rad/s unit and $F$ is the energy dependent density of states multiplied by the overlap of the vibrational states of $^3E$ and $^1A_1$ electronic states, and $\Delta$ is the energy gap between $^3E$ and $^1A_1$ states. Here, we follow the convention in Ref.~\onlinecite{DohertyGoldman2015} for the definition of $\lambda_\perp$. 
We note that Eq.~\ref{eq:ISCrate} implicitly assumes  that the electronic states of $^3E$ and $^1A_1$ participating in the ISC process do not change their character during the motion of nuclei, i.e., the $\lambda_\perp$ remains fixed independently from the coordinates of the atoms. This assumption is an analog to the Condon approximation on the optical excitation of polyatomic systems that we call basic assumption (i). This theory can explain the ISC process from $A_1$ state.

We demonstrate below that by invoking the DJT nature of $^3E$ triplet state, the three ISC rates at cryogenic temperatures \cite{Goldman:PRB2015, DohertyGoldman2015} can be naturally explained. To this end, we express the four $m_s=\pm1$ electron-phonon coupled $^3E$ wave functions in the Born-Oppenheimer basis of symmetry adapted terms, $\left|\tilde{A}_{1}\right\rangle$,  $\left|\tilde{A}_{2}\right\rangle$, $\left|\tilde{E}_{1}\right\rangle$, 
$\left|\tilde{E}_{2}\right\rangle$
(see Supplemental Materials). The symmetry adapted basis allows us to determine the phonon-induced or DJT induced mixing of electronic orbitals between the $m_s=\pm 1$ states. One can realize that degenerate $\left|\tilde{E}_{1}\right\rangle$ and $\left|\tilde{E}_{2}\right\rangle$, and $\left|\tilde{A}_{2}\right\rangle$ vibronic wave functions contain the $\left|A_{1}\right\rangle $ electronic orbital which makes the spin-orbit mediated scattering to the $^1A_1$ singlet state feasible. This explains the three different ISC rates even at cryogenic temperature. By taking the vibronic nature of
these $m_s=\pm1$ states into account, the corresponding ISC rates can be written as
\begin{equation}
\label{eq:G_A1}
\Gamma_{A_{1}}=4\pi\hbar\lambda_{\perp}^{2}\sum_{i=1}^{\infty}\left[c_{i}^{2}F\left(\Delta-n_{i}\hbar\omega_{e}\right)\right] \text{,}
\end{equation}
\begin{equation}
\label{eq:G_E12}
\Gamma_{E_{12}}=4\pi\hbar\lambda_{\perp}^{2}\sum_{i=1}^{\infty}\left[\frac{d_{i}^{2}}{2}F\left(\Delta-n_{i}\hbar\omega_{e}\right)\right] \text{,}\\
\end{equation}
\begin{equation}
\label{eq:G_A2}
\Gamma_{A_{2}}=4\pi\hbar\lambda_{\perp}^{2}\sum_{i=1}^{\infty}\left[f_{i}^{2}F\left(\Delta-n_{i}\hbar\omega_{e}\right)\right] \text{,}
\end{equation}
where $c_i$, $d_i$ and $f_i$ expansion coefficients are calculated \emph{ab initio} by solving the electron-phonon Hamiltonian, and $n_i$ is the quantum number of the phonons of $^1A_1$ electronic state.

\begin{figure}
	\includegraphics[width=1.00\columnwidth]{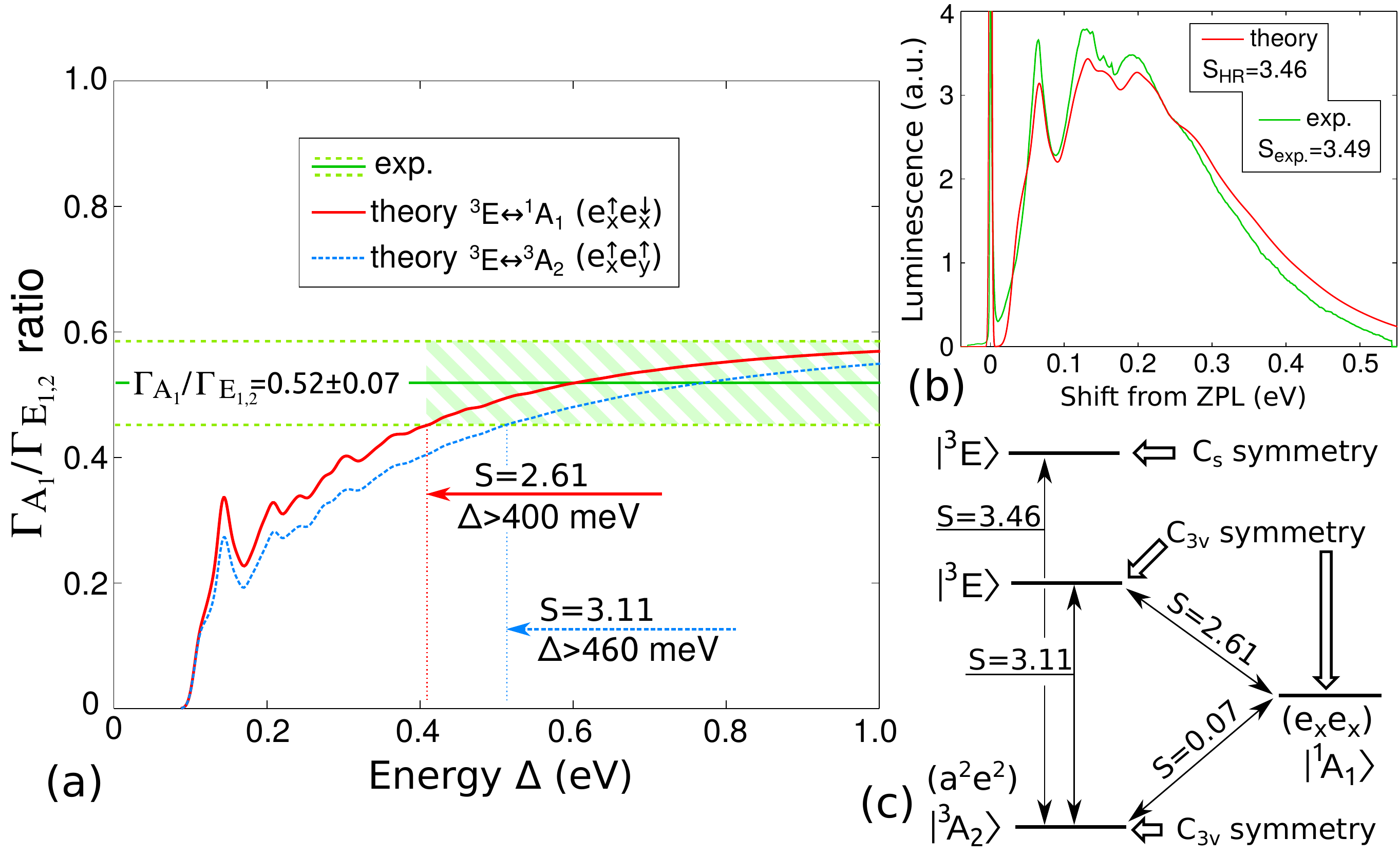}
	\caption{\label{fig:HRfit} The DJT theory of ISC rates on NV($-$) center. (a) The measured ratio of ISC rates (straight horizontal line) with the given error bar (dotted horizontal lines) from Ref.~\onlinecite{DohertyGoldman2015}) and the calculated ISC rates as a function of $\Delta$ with two $S$ parameters that are derived from two geometry changes as explained in the legend and (c) . (b) The experimental (Ref.~\onlinecite{Kehayias:2013}) and our \emph{ab initio} simulated PL spectrum where $S_\text{exp}$ and $S_\text{HR}$ are the experimentally deduced and calculated Huang-Rhys factors. The contribution of ZPL (sharp peak) to the full PL spectrum is about 4\%. (c) Diagram about the calculated $S$ factors. The energy levels are not scaled for the sake of clarity. The geometry of $^1A_1$ singlet is approximated from that of $(e_xe_x)$ singlet determinant.} 
\end{figure}

Here, we still applied the basic assumption (i) in the ISC process but we explicitly consider the DJT nature of $^3E$ state. We found that $f_i^2$ is very small (see Supplemental Materials), thus $\Gamma_{A_{2}}$ is two orders of magnitude smaller than $\Gamma_{A_{1}}$ or $\Gamma_{E_{1,2}}$, in agreement with the experiment. The ratio of $\Gamma_{E_{1,2}}$/$\Gamma_{A_{1}}$ requires the explicit calculation of $F$ function that depends on $\Delta$ and the overlap of phonon states of the electronic states. As the value of $\Delta$ is unknown we use it as a parameter. The theoretical upper limit of $\Delta$ can be calculated from the ZPL energies of the visible and near-infrared (NIR) optical transitions \cite{Kehayias:2013} that results in $\Delta < 0.75$~eV which ensures that the $^1E$ singlet level is above that of $^3A_2$ groundstate. Regarding the overlap of phonon states, we apply the Huang-Rhys approximation to calculate this quantity that is within our assumption (i) but further assumes that the parabolic APES of the electronic states so the phonon energies and states are the same in the two electronic states involved in ISC. This Huang-Rhys approximation was already implicitly employed by using $\omega_{e}$  in Eqs.~\ref{eq:G_A1}-\ref{eq:G_A2} where the value $\hbar \omega_e$ can be read in Table~\ref{tab:results}. As the ISC occurs between $^3E$ and $^1A_1$ states, the optimized geometry of these states are required in the calculation of their phonons overlap function that is characterized by its $S$ Huang-Rhys factor (see Supplementary Materials). However, Kohn-Sham HSE06 DFT cannot explicitly calculate the $^1A_1$ state. An upper bound theoretical limit on the $S$ factor can be taken from the geometry change between the $^3A_2$ groundstate and $^3E$ excited state where the DJT feature in $^3E$ geometry should be eliminated. Our calculated $S$ factor for this optical transition agrees well with the value deduced from the experimental PL spectrum of NV($-$) [see Fig.~\ref{fig:HRfit}(b)] that confirms the accuracy of our \emph{ab initio} approach \footnote{We note that Ref.~\onlinecite{Alkauskas:NJP2014} obtained $S=3.63$ for $^3A_2\rightarrow^3E$ optical transition at the same size of supercell at a C$_{3v}$ configuration that was obtained by smearing the occupation of the $e$ orbitals evenly in $^3E$ excited state.}. By eliminating the DJT feature in the phonon overlap function, we find $S=3.11$  [see Fig.~\ref{fig:HRfit}(c)].  By assuming that the geometry of the $^3A_2$ groundstate and the $^1A_1$ singlet is the same because of sharing the same $e^2$ electronic configuration, $S=3.11$ can be a theoretical upper bound limit. However, sharing the same $e^2$ electronic configuration does not guarantee the same optimized geometries for $^3A_2$, and $^1A_1$ and $^1E$ singlets. Indeed, NIR PL and absorption studies found $S=0.9$ for the $^1A_1\leftrightarrow^1E$ optical transition in NV($-$) \cite{Kehayias:2013}, thus the geometries of the two singlets differs. We conclude that it is not well supported to assume that the geometries of $^1A_1$ and $^3A_2$ is exactly the same. Therefore, we roughly approximate the geometry of $^1A_1$ state from $(e_xe_x)$ singlet spinpolarized HSE06 DFT calculation. We find $S=2.61$ with this procedure that reflects a small change in the geometry [see Fig.~\ref{fig:HRfit}(c)]. The final result on the ratio of $\Gamma_{E_{1,2}}$/$\Gamma_{A_{1}}$ as a function of $\Delta$ is plotted in Fig.~\ref{fig:HRfit}(a). As can be seen the $S=3.11$ curve implies too large $\Delta$ values going above the theoretical upper limit in a wide region when the experimental ratio is reproduced. On the other hand, $S=2.61$ curve mostly provides reasonable $\Delta$ values. This implies that $S$ should be indeed significantly smaller than that of $S=3.11$ derived from the groundstate geometry.  

In order to calculate the ISC rates, $\lambda_\perp$ should be determined. Unlike the case of $\lambda_z$, one has to apply an approximation to do this, namely, that the Kohn-Sham wave functions building up the $^3E$ state and the $^1A_1$ multiplet do not change. This is basic assumption (ii) in the ISC rate calculation which permits to compute $\lambda_\perp$ \emph{ab initio} by using the $a_1$ and $e_{x,y}$ Kohn-Sham wave functions of the NV($-$) in the $^3E$ excited state [see Fig.~\ref{fig:NV}(b)]. The converged intrinsic value is 
$\lambda_\perp = 56.3$~GHz which is relatively large, and it is not damped by DJT because $^1A_1$ state is not the part of DJT effect. We note that the nitrogen contribution is minor in $\lambda_z$ whereas it is significant in $\lambda_\perp$ that explains the large anisotropy between $\lambda_z$ and $\lambda_\perp$.  With these $\lambda_\perp$, $S$ factors and reasonable $\Delta$ values in Fig.~\ref{fig:HRfit}(a), we obtain an order of magnitude larger ISC rates than the experimental ones at $\Gamma_{A_1}=16.0$~MHz and $\Gamma_{E_{1,2}}=8.3$~MHz \cite{DohertyGoldman2015}. We conclude that $\lambda_\perp$ might be too large. We suspect that the large $\lambda_\perp$ is the consequence of the two basic assumptions, particularly, of assumption (ii). The exact determination of $\Delta$ and $\lambda_\perp$ requires an accurate \emph{ab initio} calculation of the $^1A_1$ multiplet state as a function of configuration coordinate 
of NV($-$) center. That would make possible a full \emph{ab initio} calculation of the ISC rates. Finally, we emphasize here that the ratio of the low-temperature multiple ISC rates is quantitatively reproduced by our DJT theory.


In conclusion, we demonstrated on NV center in diamond that the spin-orbit
coupling can be well
calculated from \emph{ab initio} methods in dynamic Jahn-Teller systems. We
implemented and applied a method to
calculate the damping factor on spin-orbit coupling caused by DJT effect from
first principles. Our
theory revealed that the strong coupling between electrons and phonons is responsible for the multiple scattering rates at cryogenic temperatures. Our results demonstrate the power of \emph{ab initio} modeling of this complex system that can be applied to other solid state qubits, in order to predict their key properties (electron-phonon coupling and spin-orbit coupling) that determine their initialization and readout.


Support from the EU Commission (DIADEMS project contract No.~611143) is
acknowledged.


%

\end{document}